\documentclass[aps,prd,reprint,showkeys,nofootinbib]{revtex4-1}
\usepackage{graphicx}
\usepackage{multirow}
\usepackage{amsmath}
\usepackage{slashed}
\usepackage{soul}
\usepackage{array}
\usepackage{natbib}
\begin{document}
\title{Studying the process $\gamma d \to \pi^0\eta d$}
\author{A.~Mart\'inez~Torres}
\email{amartine@if.usp.br}
\affiliation{Universidade de Sao Paulo, Instituto de Fisica, C.P. 05389-970, Sao Paulo,  Brazil.}
\affiliation{Centro Mixto Universidad de Valencia-CSIC Institutos de Investigaci\'on de Paterna, Aptdo. 22085, 46071 Valencia, Spain.}
\author{K.~P.~Khemchandani}
\affiliation{Universidade Federal de S\~ao Paulo, C.P. 01302-907, S\~ao Paulo, Brazil.}
\affiliation{Centro Mixto Universidad de Valencia-CSIC Institutos de Investigaci\'on de Paterna, Aptdo. 22085, 46071 Valencia, Spain.}
\author{E. Oset}
\affiliation{Centro Mixto Universidad de Valencia-CSIC Institutos de Investigaci\'on de Paterna, Aptdo. 22085, 46071 Valencia, Spain.}
\preprint{}

\date{\today}

\begin{abstract}
In these proceedings we present our recent results on the study of the process $\gamma d \to \pi^0 \eta d$, where the existence of a dibaryon in the $\eta d$ invariant mass distribution has been recently claimed. As we will show, many of the relevant aspects observed in the experiment, as the shift of the $\eta d$ and $\pi d$ invariant mass distributions with respect to phase space can be described with our model, where no dibaryon is formed. Instead, we consider the interaction of the $\gamma$ with the nucleons forming the deuteron to proceed through $\gamma N \to \Delta(1700)\to \eta \Delta(1232) \to  \eta \pi^0 N$, followed by the rescattering of the $\pi$ and $\eta$ with the other nucleon of the deuteron. Theoretical uncertainties related to different parameterizations of the deuteron wave function are investigated.
\end{abstract}

\maketitle
\section{Introduction}
In Refs.~\cite{Ishikawa:2021yyz,Ishikawa:2022mgt} the $\gamma d\to \pi^0\eta d$ reaction was investigated and a clear shift with respect to phase space was observed in the $\eta d$ and $\pi d$ invariant mass distributions. These distributions were fitted by considering a phenomenological model in which a dibaryon $D_{12}(2150)$ [$I(J^P)=1 (2^+)$] and a pole near the $\eta d$ threshold, with quantum numbers $I(J^P)=0 (1^-)$, are introduced in the $\pi^0 d$ and $\eta d$ invariant masses. By fitting the data, the mass and width of both dibaryons are obtained and found to be compatible with the corresponding values determined in the theoretical works claiming their existence~\cite{Gal:2014zia,Kolesnikov:2021hnv,Green:1997yia,Fix:2000hf,Barnea:2015lia}. Interestingly, the $D_{12}(2150)$ dibaryon found in Refs.~\cite{Gal:2014zia,Kolesnikov:2021hnv} has been recently explained in Ref.~\cite{Ikeno:2021frl} as a $\Delta(1232) np$ triangle singularity of the reaction $pp\to\pi^+ d$, where $pp\to\Delta^+ p$, followed by $\Delta^+\to \pi^+ n$ and the latter $n$ together with the former $p$ in the final state get bound in the form of a deuteron. Having this in mind one can question then whether the presence of a dibaryon in the $\eta d$ invariant mass distribution of the reaction $\gamma d\to\pi^0\eta d$ is needed to explain the energy dependence  observed in the experiment for the differential cross sections as a function of the $\eta d$ and $\pi^0 d$ invariant masses. This is our present topic of research.

The $\gamma d\to \pi^0\eta d$ process was theoretically investigated in Refs.~\cite{Egorov:2013ppa,Egorov:2020xdt}, with the $\eta NN$ interaction being implemented considering different sets for the scattering length of $\eta N$~\cite{Wilkin:1993fe,Green:1997yia}. However the two models produce substantial differences in the respective $\eta d$ and $\pi^0 d$ invariant mass distributions and it is not clear if the existence of a dibaryon in the $\eta NN$ system is compatible with the $\eta d$ invariant mass distribution found in Ref.~\cite{Ishikawa:2022mgt}.

The existence of a $\eta$ bound state has been a long standing puzzle~\cite{Bhalerao:1985cr,Haider:1986sa} (for a review on the $\eta N$ interaction we refer the reader to Ref.~\cite{Bass:2018xmz}). While theoretical calculations show that $\eta$ bound states can appear for medium and heavy nuclei~\cite{Chiang:1990ft,Inoue:2002xw}, their widths are quite large when compared with the corresponding binding and no definite conclusion has been drawn about the existence of such $\eta$ bound states in nuclei~\cite{Machner:2014ona,Kelkar:2013lwa,Metag:2017yuh}. Even the existence of $\eta\, {}^3$He and $\eta\,{}^4$He bound states is still unclear: while some models find a pole in the continuum, others suggest that deeper potentials than the current ones would be necessary in order to bind the $\eta$ in ${}^3$He or ${}^4$He~\cite{Xie:2016zhs,Xie:2018aeg}. In this respect, experimentally, the study of the $dd\to\eta\,{}^4$He$\to\pi^0 n\,{}^3$He, $\pi^-p\,{}^3$He reactions do not find any evidence about the existence of $\eta\,{}^4$He bound states~\cite{WASA-at-COSY:2013sya,Adlarson:2016dme}.

In view of the difficulties of finding $\eta$ bound states in heavy nuclei, it does not seem promising to find an $\eta d$ bound state in Nature. This means that some other dynamics should be responsible for the $\eta d$ and $\pi d$ invariant mass distributions found in the $\gamma d\to \pi^0\eta d$ reaction studied in Refs.~\cite{Ishikawa:2021yyz,Ishikawa:2022mgt}. As we will show in this work, the formation of $\Delta(1700)$ from $\gamma N$, together with its decay to $\eta \Delta(1232)$, are basically the main mechanisms involved in the process $\gamma d\to \pi^0\eta d$.

\section{Formalism}\label{formalism}

\begin{figure}[h!]
\includegraphics[width=0.5\textwidth]{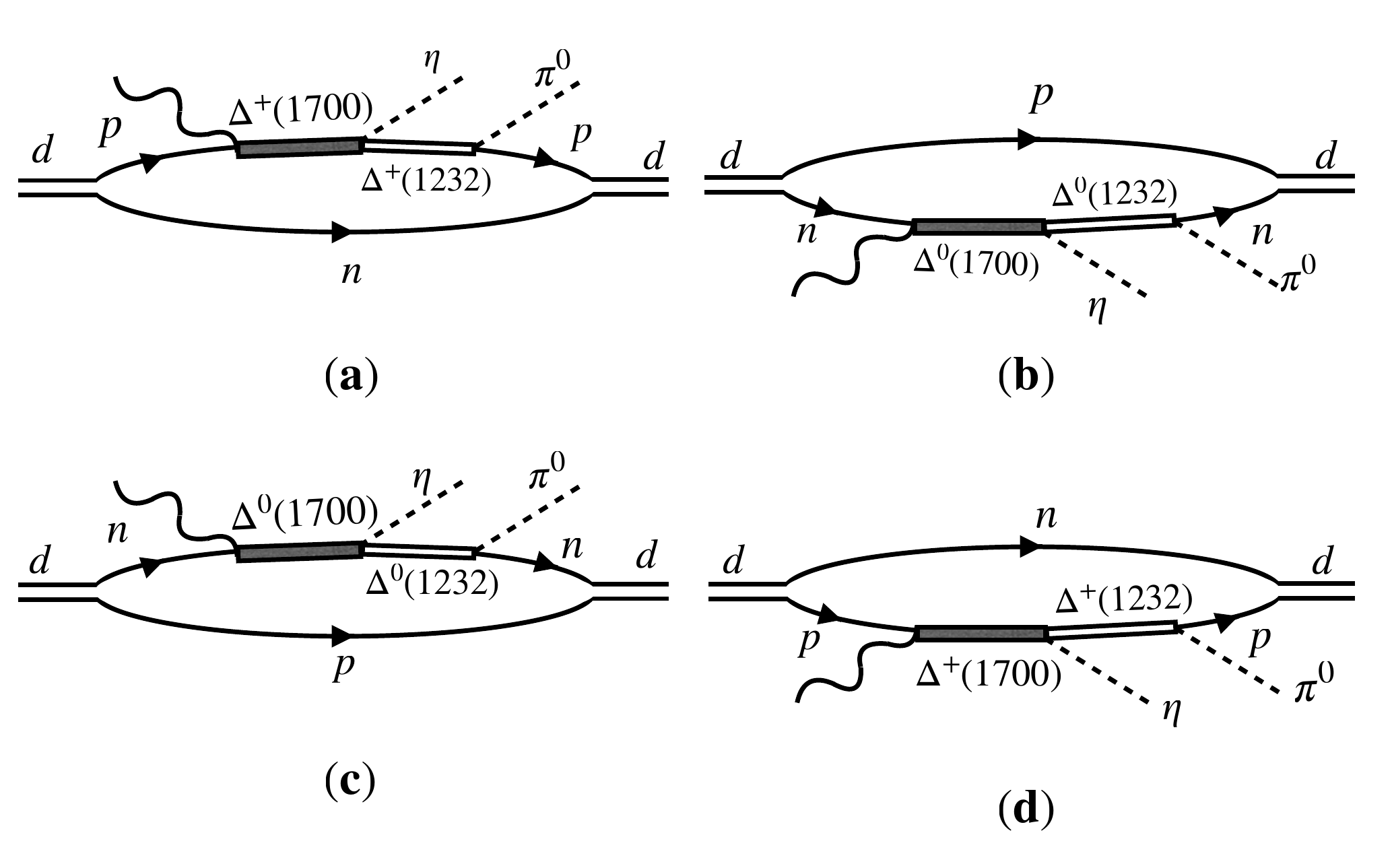} 
\caption{Different diagrams contributing to the $\gamma d \to \pi^0 \eta d$ process within the impulse approximation, i.e., without considering the rescattering of the $\pi$ and $\eta$ coming from the decay of $\Delta(1700)$ and $\Delta(1232)$, respectively.}\label{tree}
\end{figure}

In our approach, the deuteron is considered to be a $pn$ bound state with isospin $I=0$ and orbital angular momentum $L=0$, i.e.,
\begin{align}
|d\rangle=\frac{1}{\sqrt{2}}[|pn\rangle-|np\rangle].
\end{align}
In this way, to describe the process $\gamma d\to \pi^0\eta d$, we have that the photon can interact either with the $p$ and $n$ constituting the deuteron. Following Refs.~\cite{Doring:2005bx,Debastiani:2017dlz}, the interaction of a photon with a nucleon to produce a $\eta\pi^0 N$ final state proceeds through the formation of the resonance $\Delta(1700)$. This state, which was found to be generated from the dynamics involved in the s-wave interaction between pseudoscalars and baryons from the decuplet in Ref.~\cite{Sarkar:2004jh}, couples mainly to the $\eta\Delta(1232)$ channel. The $\Delta(1232)$ decays to $\pi N$, getting in this way a $\pi\eta N$ final state from $\gamma N\to\Delta(1700)\to\eta\Delta(1232)\to \eta\pi N$ (see Fig.~\ref{tree}). In the impulse approximation, i.e., without considering the rescattering of the $\eta$ and $\pi^0$, we have then four mechanisms of having $\gamma d\to\pi^0\eta d$, as shown in Fig.~\ref{tree}.

Let's determine the different contributions to the amplitude in the impulse approximation. Following Ref.~\cite{Debastiani:2017dlz}, the amplitude describing the vertex $\gamma N \to \Delta(1700)$ is given by
\begin{align}
-it_{\gamma p\Delta^*}=g_{\gamma p\Delta^*} \vec S^\dagger\cdot\vec \epsilon,\label{photo}
\end{align}
where $\Delta^*$ represents  $\Delta(1700)$, $\vec{\epsilon}$ is the polarization vector for the photon and $\vec{S}$ stands for the spin transition operator connecting states with spin 3/2 to 1/2.  The  value of the s-wave coupling $g_{\gamma p\Delta^*}$ in the preceding equation is taken to be $0.188$~\cite{Debastiani:2017dlz} (the Clebsch-Gordan coefficient $\sqrt{2/3}$ associated with the $\gamma N\to \Delta^*$ transition is already embedded on this value), which reproduces the experimental data on the radiative decay width of $\Delta(1700)$~\cite{ParticleDataGroup:2020ssz}. It is interesting to note that the amplitude in Eq.~(\ref{photo}) is the same for proton as well as neutron since the photon must behave as an isovector particle in the vertex in order to produce $\Delta(1700)$, an isospin 3/2  baryon.  

In case of the vertex $\Delta(1700)\to \Delta(1232) \eta$, we can consider
\begin{align}
-it_{\eta \Delta\Delta^*} =-i g_{\eta \Delta\Delta^*},\label{deltastrcoup}
\end{align}
with $g_{\eta \Delta\Delta^*}=1.7-i 1.4$~\cite{Sarkar:2004jh}. Finally, for the $\Delta\to \pi N$ transition, following Ref.~\cite{Ikeno:2021frl}, we have
\begin{align}
-it_{\Delta \to \pi N}=-\frac{f^*}{m_\pi}\vec S\cdot\vec p_\pi ~T^\lambda,\label{DelpiN}
\end{align}
where $\vec{p}_\pi \left(m_\pi\right)$ is the momentum (mass) of the pion, $f^*=2.13$ and $\vec S \left(T^\lambda\right)$ is the spin (isospin) transition operator acting on states with spin (isospin) $3/2$ and taking them to $1/2$. Note that the action of the isospin operator produces a factor $\sqrt{2/3}$ for  the two types of $\Delta\pi N$ vertices appearing in the diagrams in Fig.~\ref{tree}: $\Delta^+\pi^0 p$,  $\Delta^0\pi^0 n$.

In this way, we can have that
\begin{widetext}
\begin{align}
&-it_\text{impulse}=\frac{4}{\sqrt{6}}\int \frac{d^4 q}{\left(2\pi\right)^4} \left( -\frac{f^*}{m_\pi}\vec S\cdot\vec p_\pi \right) \left( g_{\gamma p\Delta^*} \vec S^\dagger\cdot\vec \epsilon \right)\left(-ig_{\eta\Delta\Delta^*}\right)\left[-i g_d\,\theta\!\left(q_{max}-|\vec p_{N}^{~d_i}|\right)\right]  \\ \nonumber
&\times\left[-i g_d\,\theta\!\left(q_{max}-|\vec p_{N}^{~d_f}|\right)\right] \frac{M_N}{E_N\left(\vec q\right)}\frac{i}{q^0-E_N\left(\vec q\right)+i\epsilon} \frac{M_N}{E_N\left(\vec p_d -\vec q\right)} ~\frac{i}{p_d^0-q^0-E_N\left(\vec p_d -\vec q\right)+i\epsilon}~ \\ \nonumber
&\times \frac{M_{\Delta^*}}{E_{\Delta^*}\left(\vec p_d -\vec q+\vec k\right)}~\frac{i}{p_d^0-q^0+k^0-E_{\Delta^*}\left(\vec p_d -\vec q+\vec k\right)+i\epsilon}~ \\ \nonumber &\times \frac{M_\Delta}{E_\Delta\left(\vec p_d -\vec q+\vec k-\vec p_\eta \right)}\frac{i}{p_d^0-q^0+k^0-p_\eta^0-E_\Delta\left(\vec p_d -\vec q+\vec k-\vec p_\eta \right)+i\epsilon} ~\\ 
&\times \frac{M_N}{E_N \left(\vec p_d -\vec q+\vec k-\vec p_\eta-\vec p_\pi \right)} \frac{i}{p_d^0-q^0+k^0-p_\eta^0-p_\pi^0-E_N\left(\vec p_d -\vec q+\vec k-\vec p_\eta -\vec p_\pi\right)+i\epsilon},\label{t1}
\end{align}
\end{widetext}
where $p_d$, $k$, $p_\eta$, and $p_\pi$ represent, respectively, the four-momentum of the deuteron in the initial state, of the initial photon and of the $\eta$ and $\pi^0$ in the final state. In Eq.~(\ref{t1}), the four-momentum $q$ is related to the nucleon inside the deuteron which does not interact with the photon of Fig.~\ref{tree}. The constant $g_d$ in Eq.~(\ref{t1}) is the $d\leftrightarrow pn$ coupling, with a value of $\left(2\pi\right)^{3/2} 2.68 \times 10^{-3}$ MeV$^{-1/2}$~\cite{Ikeno:2021frl}, while $\vec p_{N}^{~d_i} \left(\vec p_{N}^{~d_f}\right)$ represents the linear momentum of the nucleon in the rest frame of the deuteron in the initial (final) state. The $q_\text{max}$ in Eq.~(\ref{t1}) is a cut-off for the momentum of the nucleons in the rest frame of the deuteron. Within non-relativistic kinematics, which is appropriate for the process, we can write
\begin{align}\nonumber
\vec p_{N}^{~d_i}&=\frac{\vec p_d}{2}-\vec{q},\\
\vec p_{N}^{~d_f}&=\frac{\vec p_d+\vec k-\vec p_\eta-\vec p_\pi}{2}-\vec{q}.
\end{align}

Next, we can perform the $dq^0$ integration of Eq.~(\ref{t1}) by means of Cauchy's theorem, and we find

\begin{widetext}
\begin{align}\nonumber
&-it_\text{impulse}=-2i\sqrt{\frac{2}{3}}\int \frac{d^3 q}{\left(2\pi\right)^3} \left( \frac{f^*}{m_\pi}\vec S\cdot\vec p_\pi \right) \left( g_{\gamma p\Delta^*} \vec S^\dagger\cdot\vec \epsilon \right)\left(g_{\eta\Delta\Delta^*}\right)\left[g_d\,\theta\!\left(q_{max}-\biggr|\frac{\vec p_d}{2}-\vec{q}\,\biggr|\right)\right]  \\ \nonumber
&\times\left[ g_d\,\theta\!\left(q_{max}-\biggr|\frac{\vec p_d+\vec k-\vec p_\eta-\vec p_\pi}{2}-\vec{q}\,\biggr|\right)\right] \frac{M_N}{E_N\left(\vec q\right)} \frac{M_N}{E_N\left(\vec p_d -\vec q\right)} ~\frac{1}{p_d^0-E_N\left(\vec q\right)-E_N\left(\vec p_d -\vec q\right)+i\epsilon}~ \\ \nonumber
&\times \frac{M_{\Delta^*}}{E_{\Delta^*}\left(\vec p_d -\vec q+\vec k\right)}~\frac{1}{p_d^0-E_N\left(\vec q\right)+k^0-E_{\Delta^*}\left(\vec p_d -\vec q+\vec k\right)+i\epsilon}~ \\ \nonumber &\times \frac{M_\Delta}{E_\Delta\left(\vec p_d -\vec q+\vec k-\vec p_\eta \right)}\frac{1}{p_d^0-E_N\left(\vec q\right)+k^0-p_\eta^0-E_\Delta\left(\vec p_d -\vec q+\vec k-\vec p_\eta \right)+i\epsilon} ~\\ 
&\times \frac{M_N}{E_N\!\!\left(\vec p_d -\vec q+\vec k-\vec p_\eta-\vec p_\pi \!\right)} \frac{1}{p_d^0-E_N\!\left(\vec q\right)+k^0-p_\eta^0-p_\pi^0-E_N\!\left(\vec p_d -\vec q+\vec k-\vec p_\eta -\vec p_\pi\right)+i\epsilon}.\label{t2}
\end{align}
\end{widetext}
Since the reaction we are investigating involves deuterons in the initial and final states, it is convenient to introduce the corresponding wave function $\psi$ of the deuteron for a better comparison with the data. To do this, following Refs.~\cite{Ikeno:2021frl,Gamermann:2009uq}, we can replace
\begin{align}
&g_d\,\theta\!\left(q_{max}-\biggr|\frac{\vec p_d}{2}-\vec{q}\,\biggr|\right)\frac{M_N}{E_N\left(\vec q\right)}\frac{M_N}{E_N\left(\vec p_d -\vec q\right)}\nonumber\\
&\quad\times\frac{1}{p_d^0-E_N\left(\vec q\right)-E_N\left(\vec p_d -\vec q\right)+i\epsilon}\nonumber 
\end{align}
by $-\left(2\pi\right)^{3/2}\psi\left(\frac{\vec p_d}{2}-\vec{q}\right)$. Note that in our case the value $g_d$ used is compatible with the following normalization of the deuteron wave function
\begin{align}
\int d^3q |\psi(q)|^2=1.
\end{align}

Similarly, we can substitute the other combination of $g_d$, $\theta$-function and two nucleon Green's function present in the amplitude given by Eq.~(\ref{t2}) by  a $-\left(2\pi\right)^{3/2}\psi\left(\frac{\vec p_d+\vec k-\vec p_\eta-\vec p_\pi}{2}-\vec{q}\right)$. In this way, we can rewrite Eq.~(\ref{t2}) as
\begin{widetext}
\begin{align}\nonumber
&t_{impulse}=2\sqrt{\frac{2}{3}}g_{\gamma p\Delta^*}g_{\eta\Delta\Delta^*}\frac{f^*}{m_\pi}M_\Delta M_{\Delta^*} \int \frac{d^3 q}{\left(2\pi\right)^3}  \frac{(\vec S\cdot\vec p_\pi)(\vec S^\dagger\cdot\vec \epsilon)}{\left[E_{\Delta^*}\left(\vec p_d -\vec q+\vec k\right)\right]\left[E_\Delta\left(\vec p_d -\vec q+\vec k-\vec p_\eta \right)\right]}\\ \nonumber
&\times \frac{1}{p_d^0-E_N\left(\vec q\right)+k^0-E_{\Delta^*}\left(\vec p_d -\vec q+\vec k\right)+i\epsilon}~ \frac{1}{p_d^0-E_N\left(\vec q\right)+k^0-p_\eta^0-E_\Delta\left(\vec p_d -\vec q+\vec k-\vec p_\eta \right)+i\epsilon}\\
&\times \left(2\pi\right)^{3}\psi\left(\frac{\vec p_d}{2}-\vec{q}\right) \psi\left(\frac{\vec p_d+\vec k-\vec p_\eta-\vec p_\pi}{2}-\vec{q}\right).\label{t3}
\end{align}
\end{widetext}
To estimate theoretical uncertainties, we will use different well known parameterizations for the deuteron wave function, such as those of Refs.~\cite{Machleidt:2000ge,Lacombe:1981eg,Reid:1968sq,Adler:1975ga}. 

Next, Eq.~(\ref{t3}) has the spin structure $(\vec S\cdot\vec p_\pi)(\vec S^\dagger\cdot\vec \epsilon)$ and we need to evaluate the different spin transitions considering the two possible polarizations of the photon and the different spin projections of the deuteron. To do this, first, we choose the photon momentum to be parallel to the $z$-axis, such that $\vec k =\left(0,~0,~|\vec k|\right)$. In this way, the polarization vectors of the photon are given by $\vec{\epsilon}_1= \left(1,~0,~0\right)$, $\vec{\epsilon}_2=\left(0,~1,~0\right)$. Then, we make use of the useful property $\sum\limits_\text{polar.} S_i S_j^\dagger=\frac{2}{3}\delta_{ij}-\frac{i}{3}\epsilon_{ijk}\sigma_k$
and the fact that $\Delta$ is produced at the vertex $\Delta^*\to \Delta \eta$, which implies that the spin projections of $\Delta^*$ and  $\Delta$ always coincide, i.e., $m_{\Delta^*}=m_\Delta$. Then we can write
\begin{align}\nonumber
(\vec S\cdot\vec p_\pi)(\vec S^\dagger\cdot\vec \epsilon)&= \sum_{m_\Delta}~p_{\pi_i}~\epsilon_j ~S_i\mid m_\Delta\rangle\langle m_\Delta\mid S_j^\dagger\\
&=\frac{2}{3} \vec p_\pi\cdot\vec \epsilon-\frac{i}{3}\epsilon_{ijk}~p_{\pi_i}\epsilon_j\sigma_k.~\label{wij}
\end{align}
By means of Eq.~(\ref{wij}), we can now evaluate the corresponding matrix elements for the different polarization vectors of the incident photon and the spin projections of the nucleons forming the deuteron. Let us denote these matrix elements by $W_{\mu,\mu^\prime}^\lambda$, where the indices $\mu, \mu^\prime=-1, 0, 1$ represent, respectively, the spin projections $\downarrow\downarrow$, $\uparrow\downarrow+\downarrow\uparrow$, and  $\uparrow\uparrow$ of the nucleons in the deuteron, and $\lambda=1,2$ represents the two possible polarization vectors of the photon.  In this way, we have, for example,
\begin{align}\nonumber
W_{1,1}^\lambda=&\langle \uparrow\uparrow|\vec S\cdot\vec p_\pi\vec S^\dagger\cdot\vec \epsilon_\lambda~ | \uparrow\uparrow\rangle,\\\nonumber
W_{1,0}^\lambda=&\langle \frac{1}{\sqrt{2}}\left(\uparrow\downarrow+\downarrow\uparrow\right)|\vec S\cdot\vec p_\pi \vec S^\dagger\cdot\vec \epsilon_\lambda| \uparrow\uparrow\rangle,\\\nonumber
W_{1,-1}^\lambda=&\langle \downarrow\downarrow|\vec S\cdot\vec p_\pi \vec S^\dagger\cdot\vec \epsilon_\lambda| \uparrow\uparrow\rangle.
\end{align}
Thus, the amplitude in the impulse approximation obtained in Eq.~(\ref{t3}) depends on the spin projections of the deuteron in the initial ($\mu$) and final ($\mu^\prime$) states, as well as on the transverse polarization of the photon ($\lambda$). It is then convenient to use the notation  $t^\lambda_{\mu,\mu^\prime}$. In Table~\ref{Tabapp1} we list all $W_{\mu,\mu^\prime}^\lambda$ matrix elements obtained in the impulse approximation.

\begin{table}[htbp!]
\caption{Spin transition elements $W_{\mu,\mu^\prime}^\lambda$. The subscripts $\mu$ and $\mu^\prime$ are related to the polarizations of the deuteron in the initial and final states. Note that $W_{\mu^\prime,\mu}^\lambda$ is the negative of the complex conjugate of $W_{\mu,\mu^\prime}^\lambda$, thus, it suffices to list any one of them.}\label{Tabapp1}
\begin{tabular}{ccc}
\hline\hline
$\mu$& $\mu^\prime$&$W_{\mu,\mu^\prime}^\lambda$\\\hline
1&1&$\frac{2}{3} \vec p_\pi\cdot\vec \epsilon_\lambda-\frac{i}{3}\left(p_{\pi_x}\epsilon_{\lambda_y}-p_{\pi_y}\epsilon_{\lambda_x}\right)$ \\
1&0&$-\frac{i}{3\sqrt{2}}\left(-p_{\pi_z}\epsilon_{\lambda_y}+ip_{\pi_z}\epsilon_{\lambda_x}\right)$\\
1&-1&0\\
0&0&$\frac{2}{3} \vec p_\pi\cdot\vec \epsilon_\lambda$\\
0&-1&$-\frac{i}{3\sqrt{2}}\left(-p_{\pi_z}\epsilon_{\lambda_y}+ip_{\pi_z}\epsilon_{\lambda_x}\right)$\\
-1&-1&$\frac{2}{3} \vec p_\pi\cdot\vec \epsilon_\lambda+\frac{i}{3}\left(p_{\pi_x}\epsilon_{\lambda_y}-p_{\pi_y}\epsilon_{\lambda_x}\right)$\\\hline\hline
\end{tabular}
\end{table}

\begin{figure}[h!]
\begin{tabular}{ll}
\includegraphics[width=0.2\textwidth]{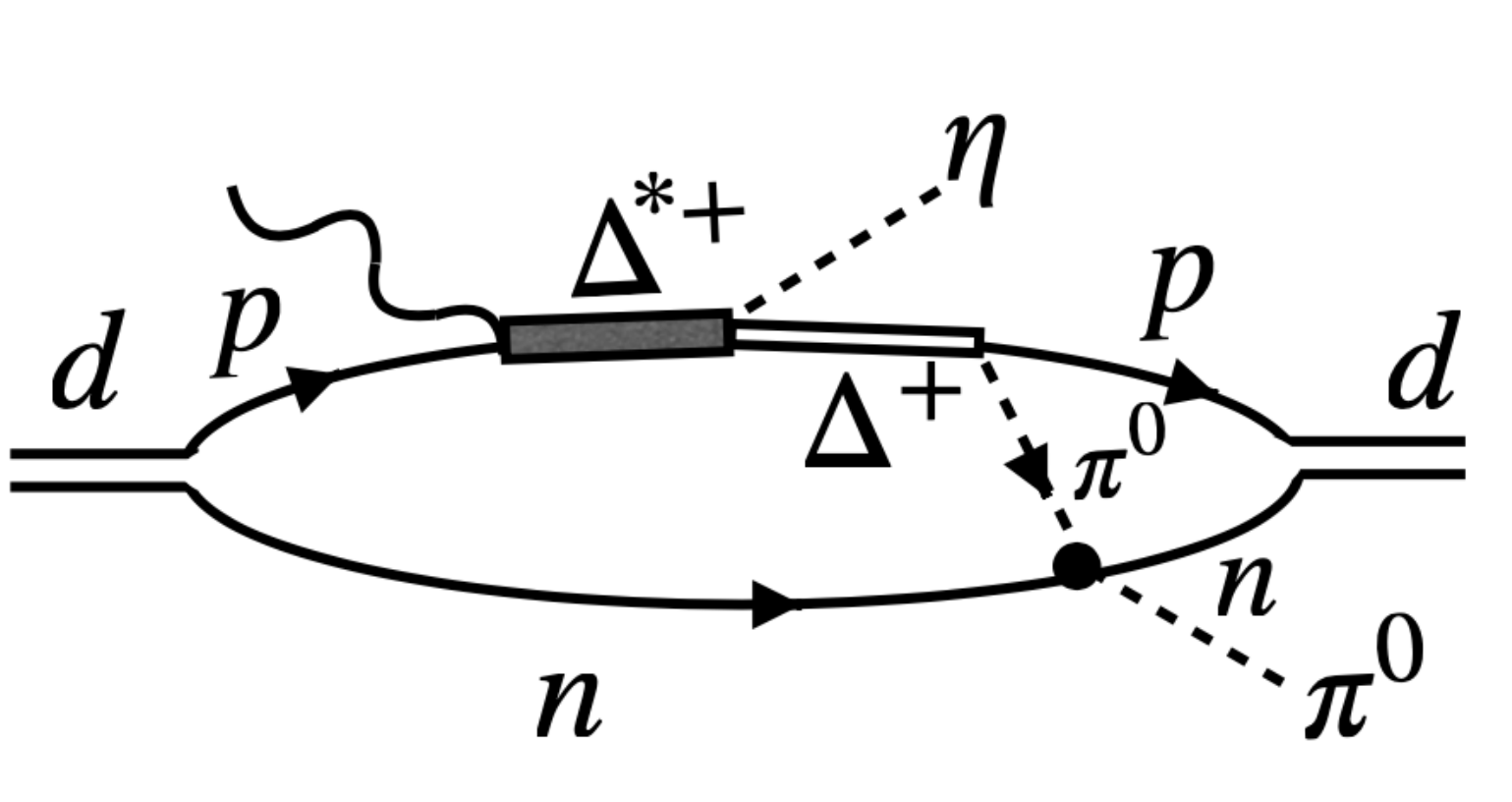}&\includegraphics[width=0.2\textwidth]{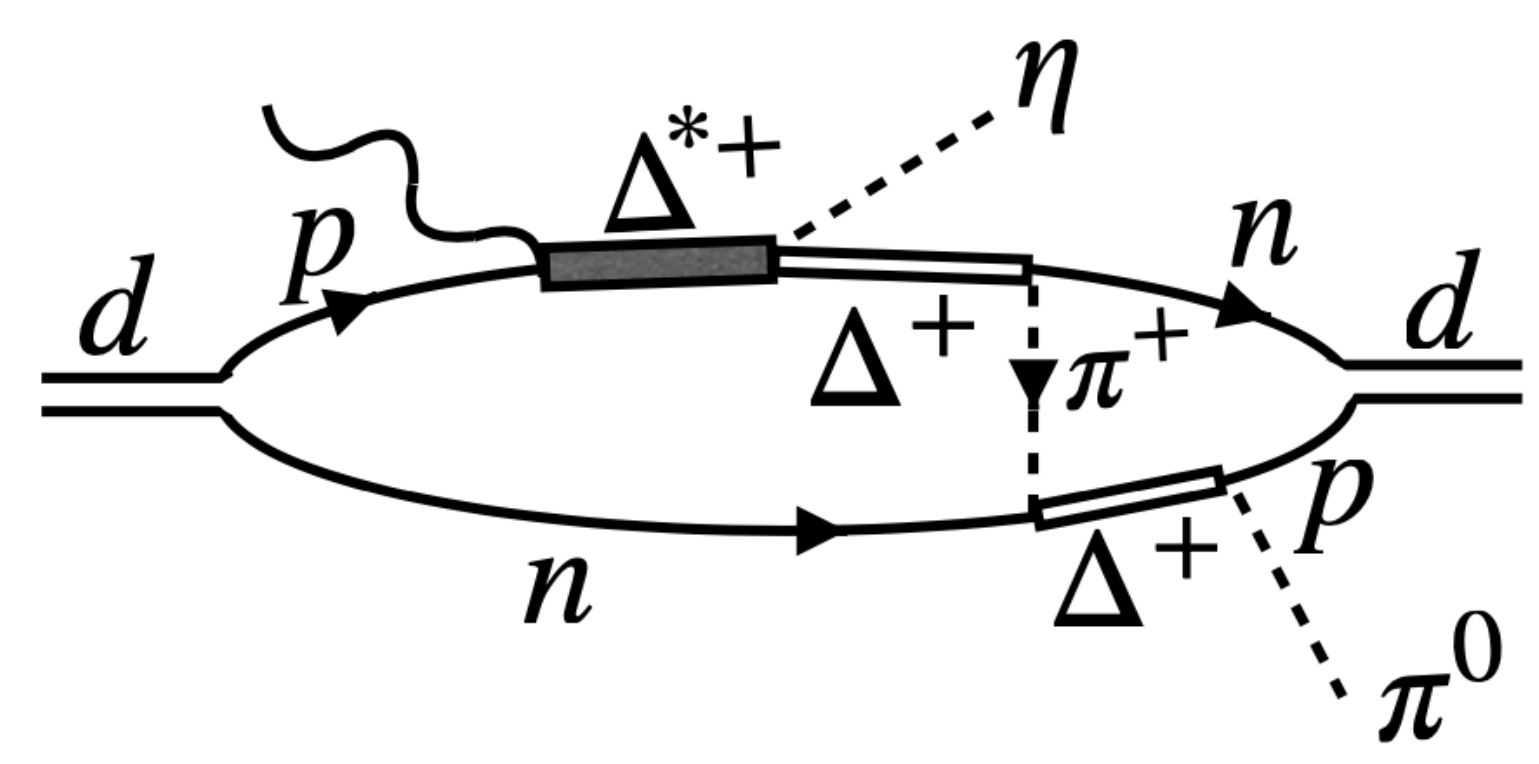}
\end{tabular}
\caption{Some of the diagrams contributing to the rescattering of the pion in the intermediate state in s- (left) and p-waves (right). The thick dot stands for the s-wave $\pi N \to \pi N$ interaction. There are eight diagrams in total involving the rescattering of a pion in s-wave and another eight diagrams for the rescattering of a pion in p-wave. For the full set of diagrams we refer the reader to Ref.~\cite{MartinezTorres:2022evx}.}\label{piresc}
\end{figure}

After evaluating the amplitude in the impulse approximation, the next contribution to the process $\gamma d\to\pi^0\eta d$ in our approach corresponds to the rescattering of $\pi$ and $\eta$. The $\pi$ can rescatter with one of the nucleons of the deuteron in s-wave as well as in p-wave, i.e., relative orbital angular momentum $l=0,1$ (we illustrate some of the corresponding diagrams in Fig.~\ref{piresc}). In the former case, we follow the approach of Ref.~\cite{Oset:1985wt} to determine the $\pi N\to \pi N$ amplitude in s-wave, while in the latter case, the $\Delta(1232)$ is exchanged, with the $\Delta\to\pi N$ vertex being described by the amplitude in Eq.~(\ref{DelpiN}).

\begin{figure}[h!]
\includegraphics[width=0.5\textwidth]{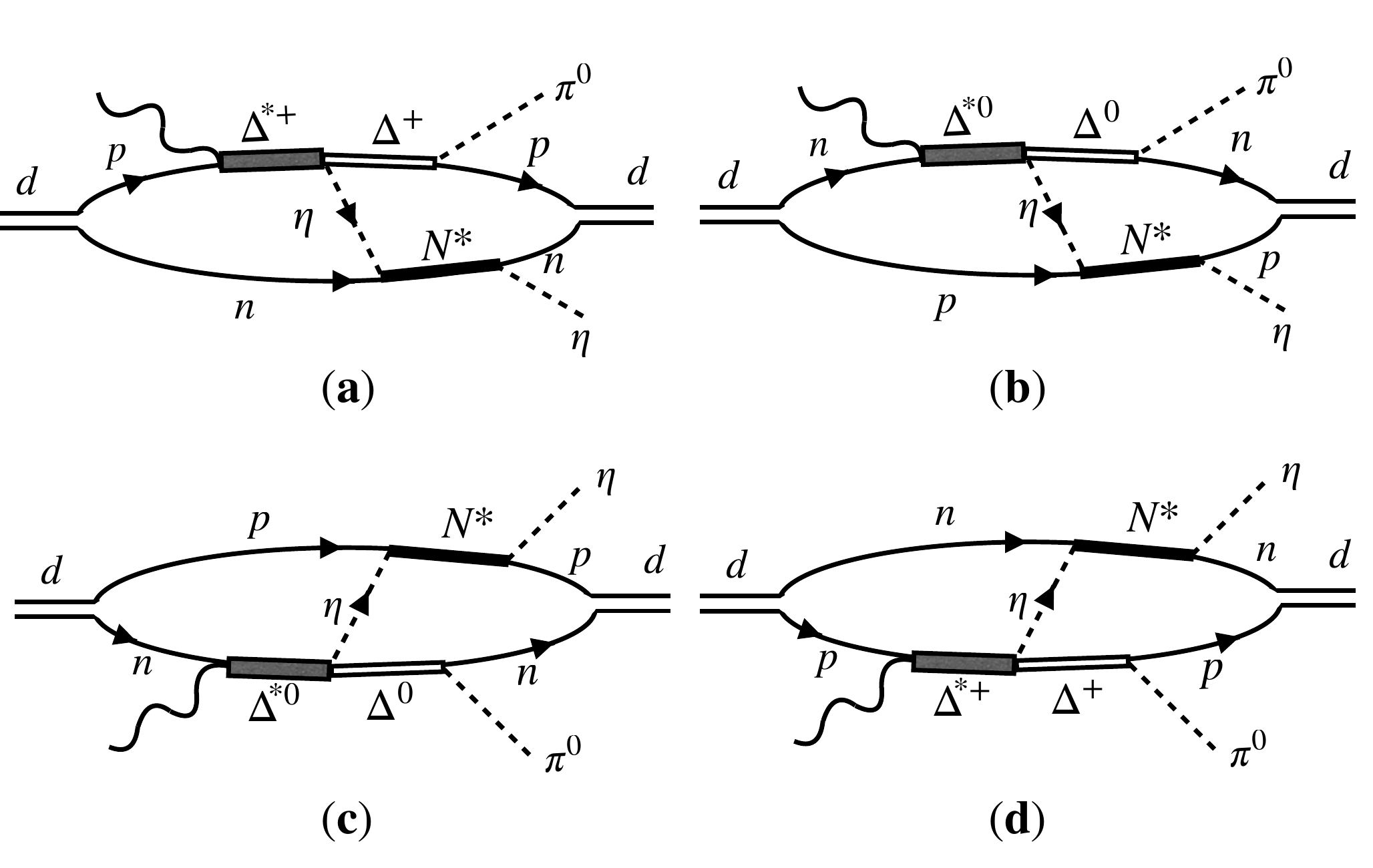}
\caption{Diagrams contributing to the s-wave rescattering of an $\eta$ with one of the nucleons of the deuteron.}\label{etares}
\end{figure}

In case of the rescattering of the $\eta$ with one of the nucleons of the deuteron, the $N^*(1535)$ can be generated in s-wave. As shown in Ref.~\cite{Inoue:2001ip}, this latter state couples mainly to $K\Sigma$ and $\eta N$, with the coupling $g_{\eta NN^*(1535)}=1.46- i 0.43$, and we have the contributions shown in Fig.~\ref{etares}.

Following the same methodology to get the contribution in the impulse approximation, we can determine the amplitudes for the rescattering of a pion in s- and p-waves as well as that  related to the rescattering of an $\eta$ in s-wave. For the explicit expressions as well as for more details on the calculations, we refer the reader to Ref.~\cite{MartinezTorres:2022evx}.

Finally, we implement in our approach the unstable nature of states like $\Delta^*(1700)$, $\Delta(1232)$ and $N^*(1535)$ by replacing $E_R-i\epsilon$ by $E_R-i\Gamma_R/2$ in the different amplitudes, where $R$ stands for a resonance/unstable state. In case of $\Delta(1232)$, we consider an energy dependent width
\begin{align}
\Gamma_\Delta\left(M_{\Delta inv}\right)=\Gamma_\Delta \frac{M_\Delta}{M_{\Delta inv}}\left(\frac{q_\pi}{q_{\pi on}}\right)^3,\label{deltawidth}
\end{align}
where $q_\pi$ and $q_{\pi on}$ are defined as
\begin{align}\nonumber
q_\pi&=\frac{\lambda^{1/2}\left(M_{\Delta inv}^2, M_N^2, m_\pi^2\right)}{2 M_{\Delta inv}},\\\nonumber
q_{\pi on}&=\frac{\lambda^{1/2}\left(M_{\Delta}^2, M_N^2, m_\pi^2\right)}{2 M_{\Delta}},
\end{align}
with
\begin{align}
M_{\Delta inv}^2= E_\Delta^2-|\vec p_\Delta |^2.
\end{align}

\section{Results and discussions}
Using the amplitudes discussed in the previous section we can determine the invariant mass distributions for $\eta d$ and $\pi^0 d$ in the final state as
\begin{align}
\frac{d\sigma}{dM_{\eta d}}&=\frac{M_d^2}{8 \big|\vec k\big| s}\frac{1}{\left(2\pi\right)^4} \big|\vec p_\pi\big|\big|\vec p^{\,R\eta d}_\eta\big|\int d cos\theta_\pi \int d\Omega^{R\eta d}_\eta\nonumber\\
&\times \overline {\sum\limits_{\mu,\lambda}} \sum_{\mu^\prime} \big| t^\lambda_{\mu,\mu^\prime}\big|^2,\nonumber\\
\frac{d\sigma}{dM_{\pi^0 d}}&=\frac{M_d^2}{8 \big|\vec k\big| s}\frac{1}{\left(2\pi\right)^4} \big|\vec p_\eta\big|\big|\vec p^{\,R\pi d}_\pi\big|\int d cos\theta_\eta \int d\Omega^{R\pi d}_\pi\nonumber\\
&\times \overline {\sum\limits_{\mu,\lambda}} \sum_{\mu^\prime} \big| t^\lambda_{\mu,\mu^\prime}\big|^2,\label{invmass2}
\end{align}
where the summation signs represent the sum over the polarizations of the particles in the initial and final states, with the bar over the sign indicating averaging over the initial state polarizations. In Eq.~(\ref{invmass2}), $s$ is the standard Mandelstam variable, $\vec p_\pi \left(\vec p_\eta\right)$ is the pion (eta) momentum in the global center of mass frame, and $\vec p^{\,R\eta d}_\eta \left(\vec p^{\,R\pi d}_\pi\right)$ denotes the eta (pion) momentum in the rest frame of $\eta d \left(\pi d\right)$. The variable $\Omega^{R\eta d}_\eta~\left(\Omega^{R\pi d}_\pi\right)$ in Eq.~(\ref{invmass2}) are the solid angle of $\eta~\left(\pi\right)$ in the $\eta d \left(\pi d\right)$ rest frame. Note that we calculate the amplitudes in the global center of mass frame, i.e., $\vec p_d +\vec k=0$ and $p_d^0+k^0$ is taken as $\sqrt{s}$. Thus, we must boost $\vec p^{\,R\pi d}_\pi$ and $\vec p^{\,R\eta d}_\eta$ to the global center of mass frame. The expressions for the boosted $\eta$ and $\vec{p}_\pi$  momenta can be found in Ref.~\cite{MartinezTorres:2022evx}.

\begin{figure}[h!]
\begin{tabular}{cc}
\includegraphics[width=0.22\textwidth]{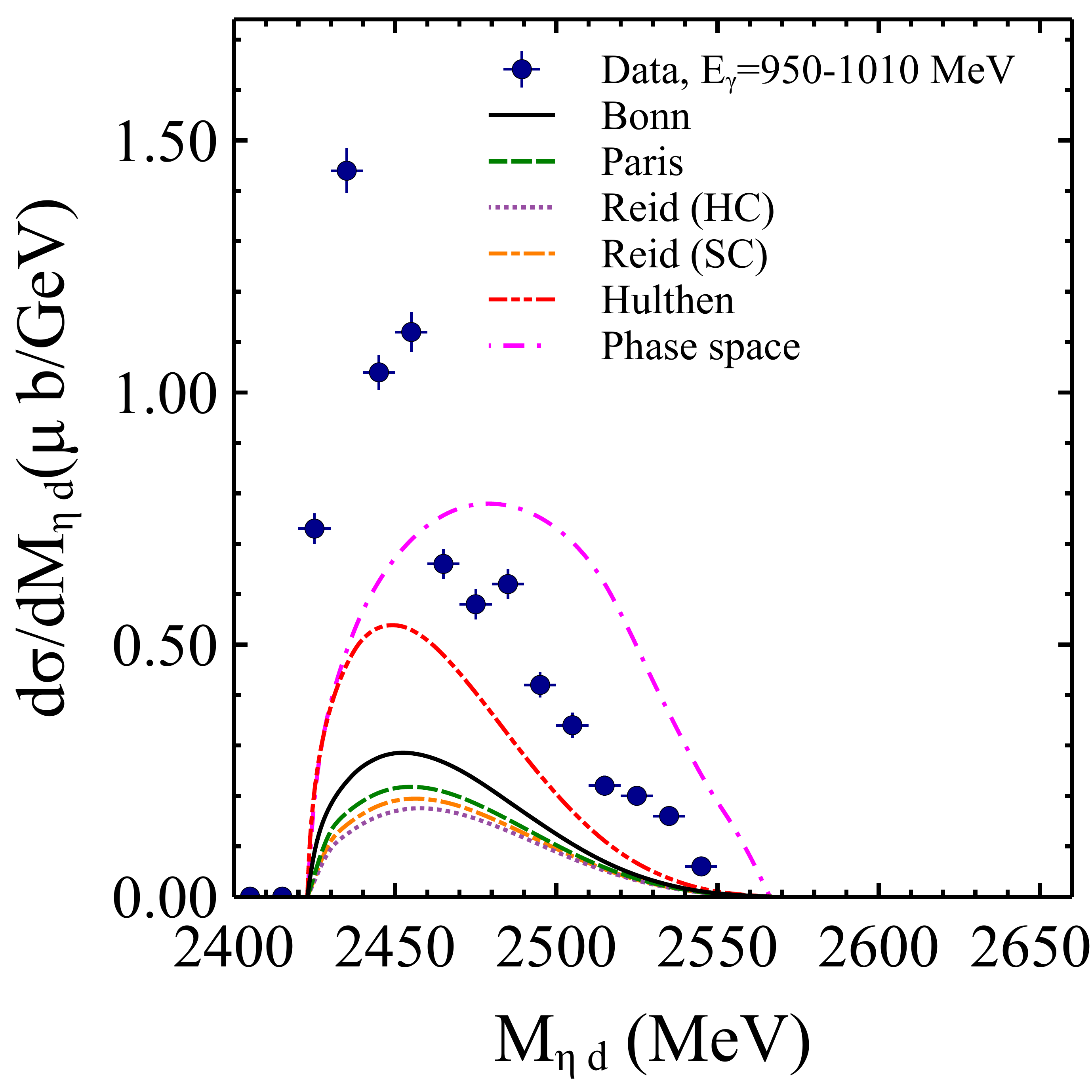} &\includegraphics[width=0.22\textwidth]{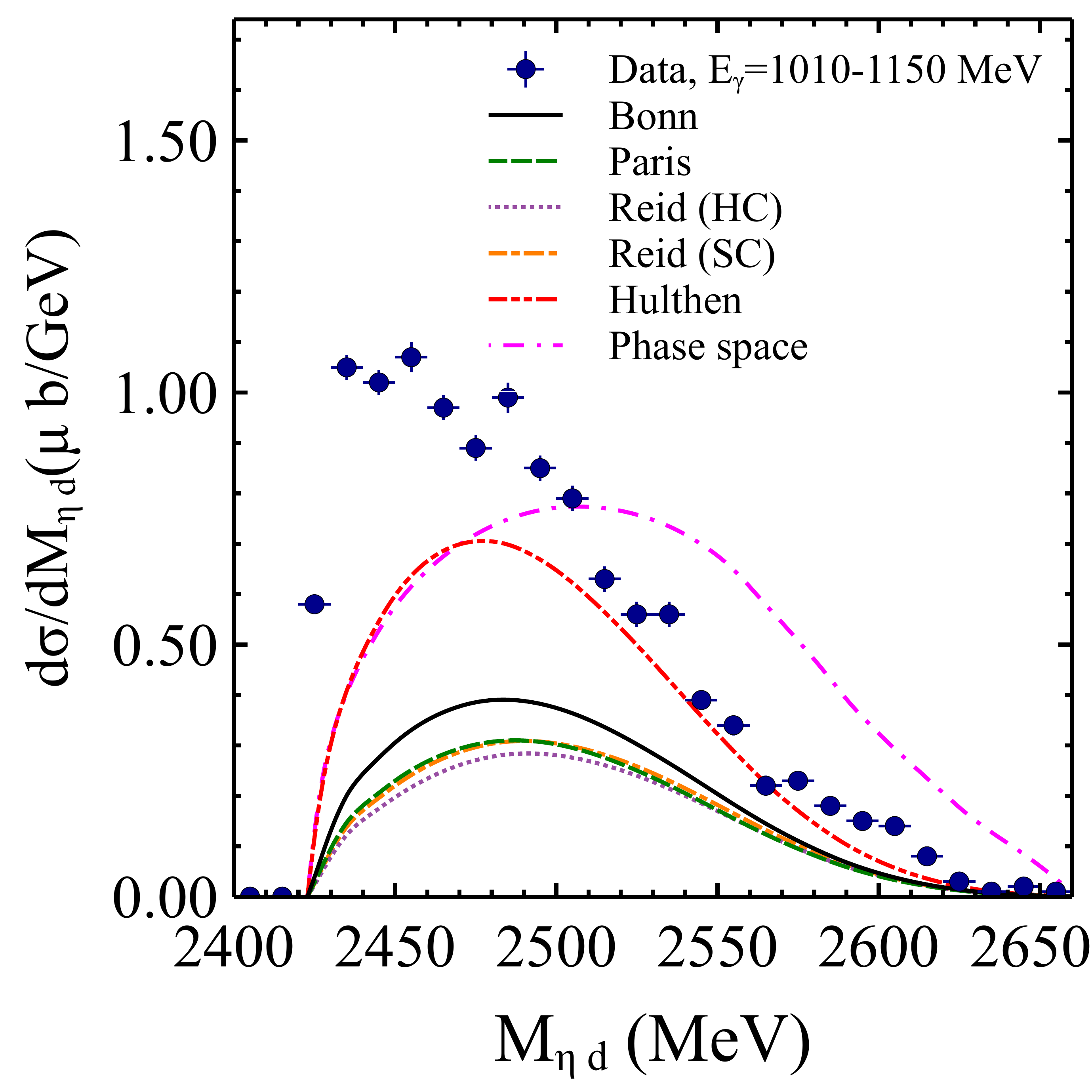}\\
\includegraphics[width=0.22\textwidth]{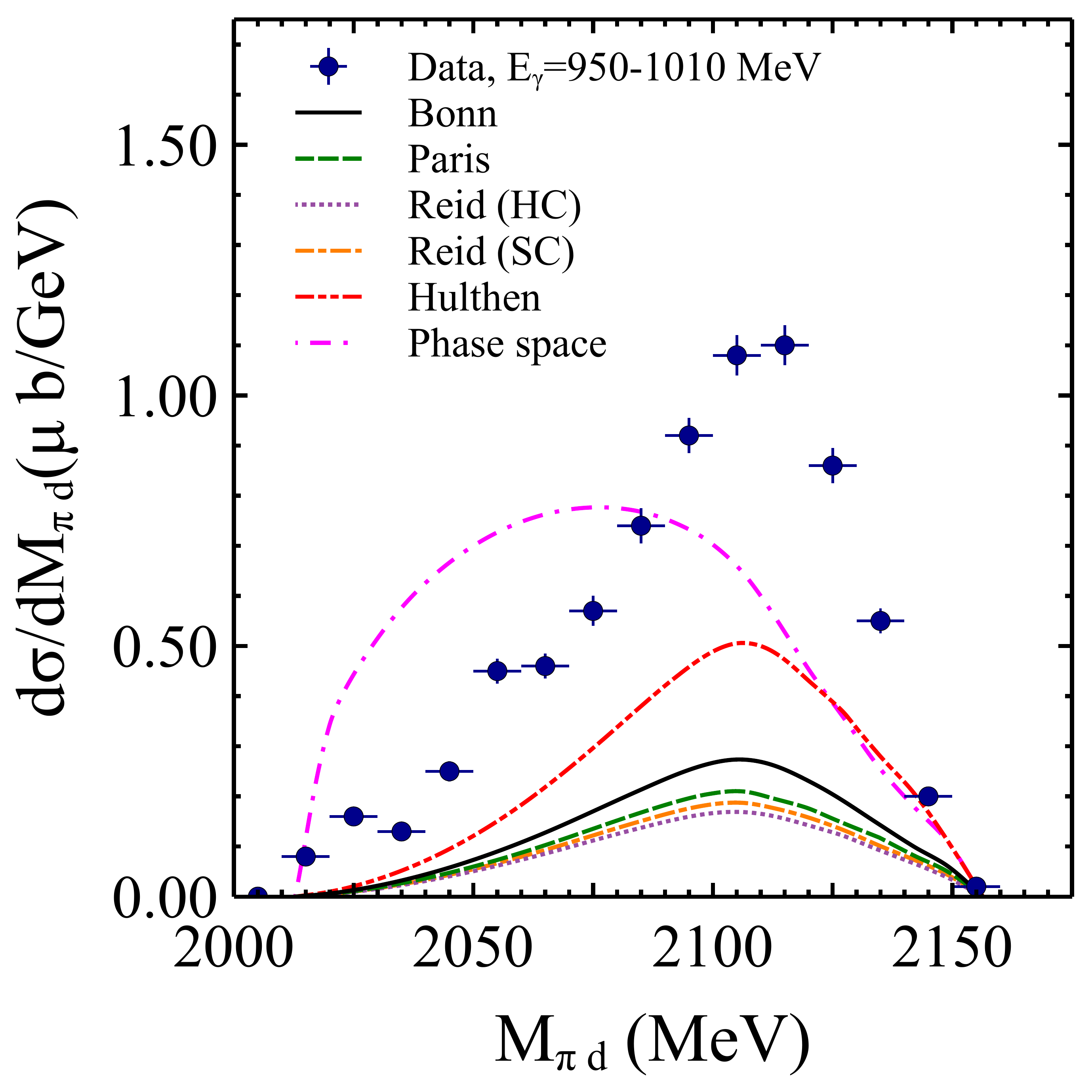} &\includegraphics[width=0.22\textwidth]{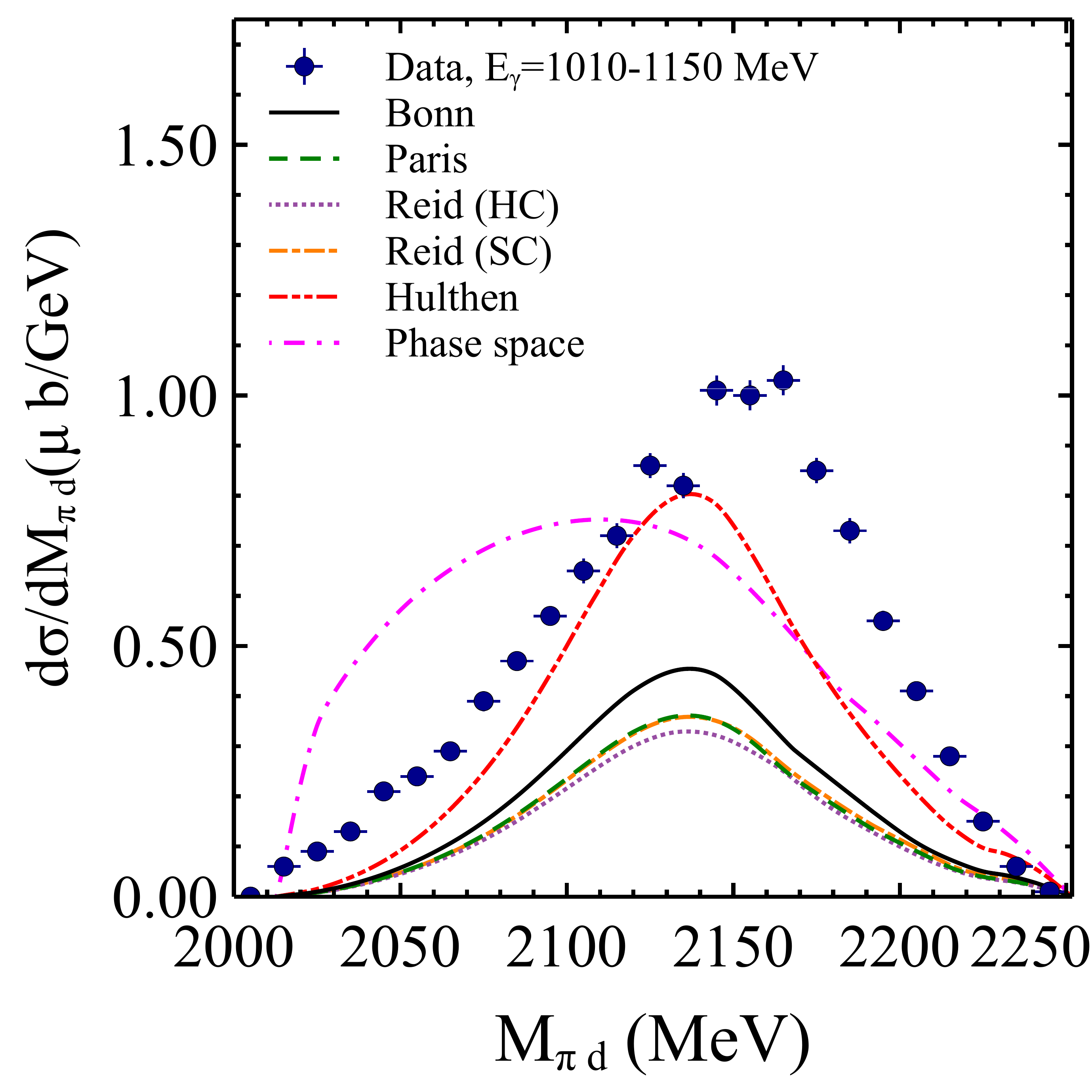}
\end{tabular}
\caption{Differential cross sections obtained in the impulse approximation as a function of the $\eta d$ (upper panels) and $\pi^0 d$ (lower panels) invariant masses. The left (right) side figures show average cross sections for the beam energy range $E_\gamma=950-1010$ MeV ($E_\gamma=950-1010$ MeV). Experimental data, shown as filled circles, are taken from Ref.~\cite{Ishikawa:2021yyz}. The deuteron wave functions considered in the calculations are based on the following parameterizations for the $NN$ potentials: Bonn~\cite{Machleidt:2000ge}, Paris~\cite{Lacombe:1981eg}, Reidt Hard-Core (HC) and Soft-Core (SC)~\cite{Reid:1968sq} and Hulth\'en~\cite{Adler:1975ga}.}\label{sigmatree}
\end{figure}

Let us discuss now the results obtained. We start with the invariant mass distributions found considering the impulse approximation. The results obtained are shown in Fig.~\ref{sigmatree}. In the experiment two different sets of photon beam energies are considered, 950-1010~MeV and 1010-1150~MeV. Thus, to compare with the experimental data we need to calculate the $\eta d$ and $\pi^0 d$ mass distributions for different energies between  950-1010~MeV as well as between 1010-1150~MeV and determine the average values. In particular, we consider the energies 950, 980 and 1010 MeV for the first energy range and 1010, 1050, 1100, and 1150 MeV for the second energy range. In Fig.~\ref{sigmatree} the left (right) panel represents the results obtained averaging the curves found for $E_\gamma=950-1010$ MeV ($E_\gamma=1010-1150$ MeV). The top (bottom) panel in Fig.~\ref{sigmatree} is related to the results obtained for the $\eta d$ ($\pi^0d$) invariant mass distribution. The different lines shown in Fig.~\ref{sigmatree} correspond to the results obtained with different parameterizations of the deuteron wave function. As can be seen, the shift with respect to phase space shown by the data of Ref.~\cite{Ishikawa:2021yyz} on the differential cross section can be reproduced with the impulse approximation, and it is a consequence of the dynamics considered (see Fig.~\ref{tree}). Indeed, the mechanism in Fig.~\ref{tree} favors the $\pi^0$ to go with as high energy as possible to place the $\Delta(1232)$ on-shell. This leaves less energy for the $\eta$ and the $\eta d$ invariant mass becomes smaller. Conversely, the $\pi^0$ goes out with larger energy than expected from phase space leading to a $\pi d$ invariant mass bigger than the phase space contribution.

\begin{figure}[h!]
\includegraphics[width=0.35\textwidth]{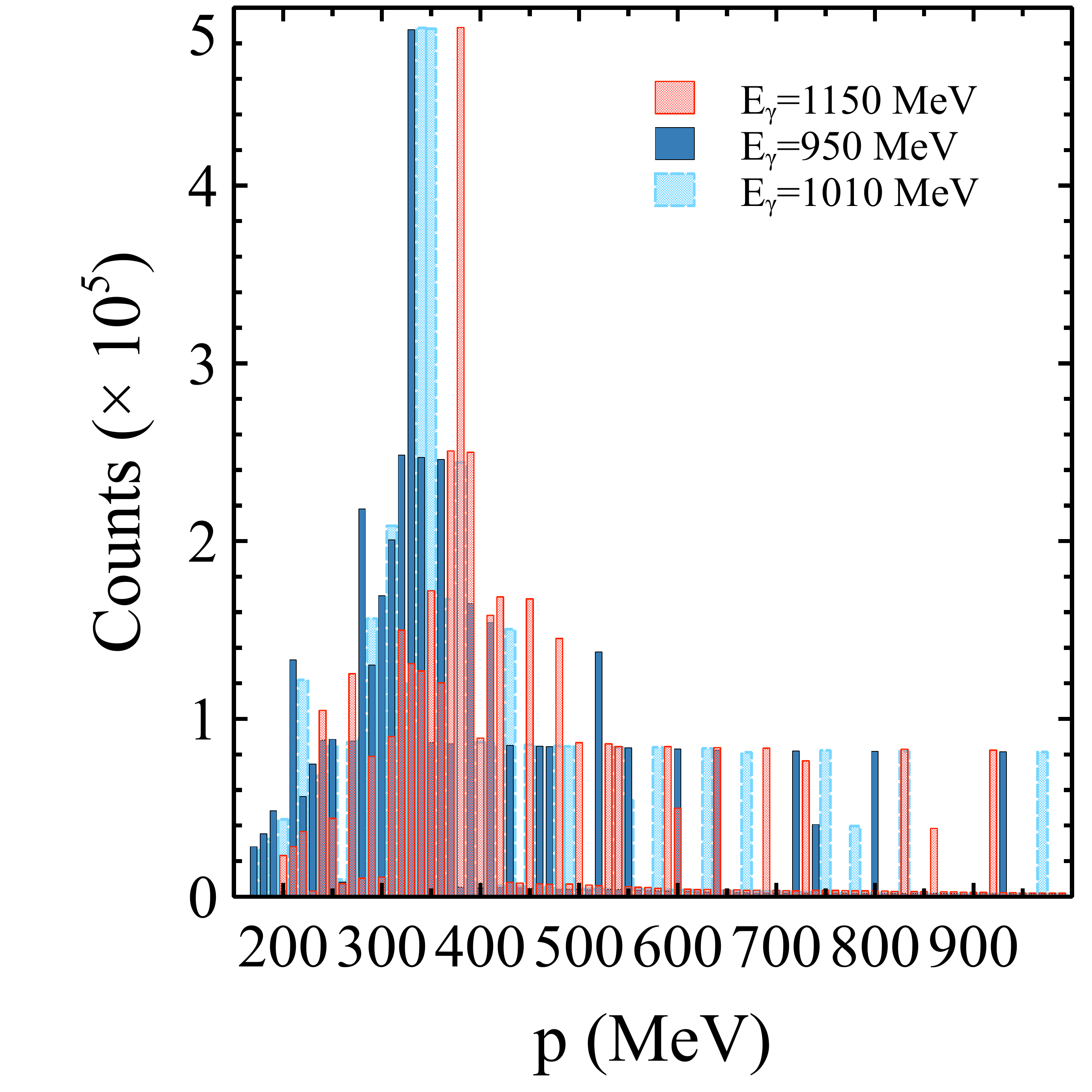}
\caption{Accumulation of events satisfying the condition in Eq.~(\ref{condition}) for values of $q_\text{max}$ in the range $0-1000$ MeV.}\label{distri}
\end{figure}

Note, however, that the magnitude of the distributions shown in Fig.~\ref{sigmatree} is substantially affected by the choice of the wave function parameterization considered in the calculations. Such differences are related to the typical momentum values of the deuteron in the reaction considered. Indeed, as can be seen in Fig.~\ref{distri}, the deuteron wave function gets determined, most frequently, in the momentum range 300-400 MeV for different values of the photon energy. 
This result has been found by generating random numbers when calculating the phase-space integration for the differential cross sections.  Then, by collecting the events which satisfy the condition 
\begin{align}
&\theta\!\left(q_{max}-\biggr|\frac{\vec p_d}{2}-\vec{q}\,\biggr|\right)\nonumber\\
&\times\theta\!\left(q_{max}-\biggr|\frac{\vec p_d+\vec k-\vec p_\eta-\vec p_\pi}{2}-\vec{q}\,\biggr|\right)
=1,\label{condition}
\end{align}
while changing $q_\text{max}$ from $10$ to $1000$ MeV, in steps of 10 MeV, we can define $R_i$ as the number found for the $i$th value of $q_\text{max}$. In this way, the difference $R_{i+1}-R_i$ provides the fraction of events where either $|\vec p_d/2-\vec{q}|$ or $|(\vec p_d+\vec k-\vec p_\eta-\vec p_\pi)/2-\vec{q}|$ are between $q_\text{max}$ and $q_\text{max}+10$ MeV. 

\begin{figure}[h!]
\includegraphics[width=0.35\textwidth]{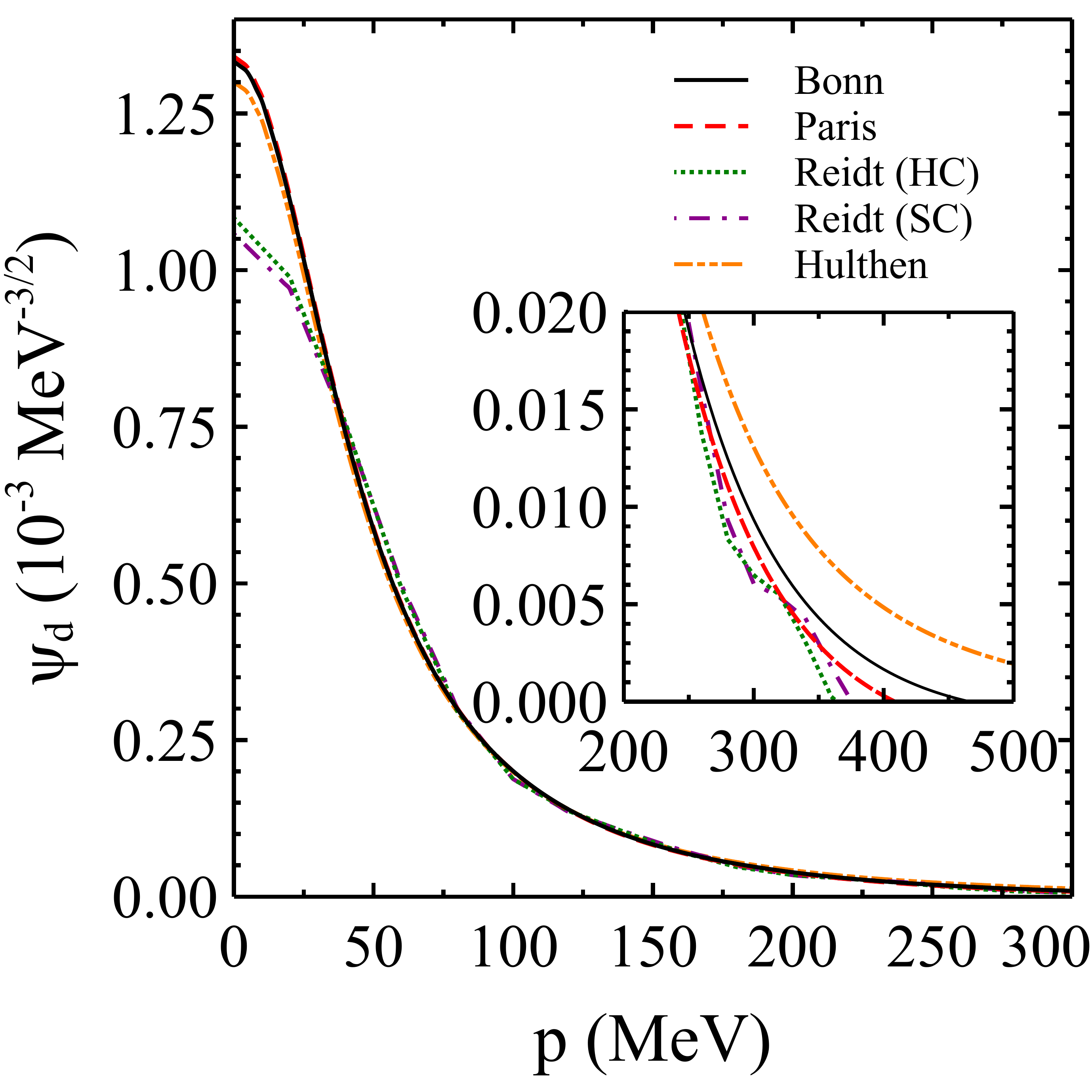}
\caption{Deuteron wave functions based on the following parameterizations for the $NN$ potentials: Bonn~\cite{Machleidt:2000ge}, Paris~\cite{Lacombe:1981eg}, Reidt Hard-Core (HC) and Soft-Core (SC)~\cite{Reid:1968sq} and Hulth\'en~\cite{Adler:1975ga}.}\label{zoom}
\end{figure}

Considering then the momentum region 300-400 MeV, as can be seen in Fig.~\ref{zoom}, different parameterizations of the deuteron wave functions produce substantial differences precisely in this momentum region. 
The different wave functions of the deuteron are related to different parameterizations of the $NN$ potential, parameterizations which are based on meson exchange potentials. Therefore, they should be expected to work at distances where the nucleons do not overlap. However, this should not be the case for momentum of the deuteron ranging between 300-400 MeV. Then the $NN$ scattering models of Refs.~\cite{Machleidt:2000ge,Lacombe:1981eg,Reid:1968sq,Adler:1975ga} cannot provide precise descriptions for the deuteron wave function in the momentum range needed to study the reaction $\gamma d\to \pi^0\eta d$.

\begin{figure}[h!]
\includegraphics[width=0.48\textwidth]{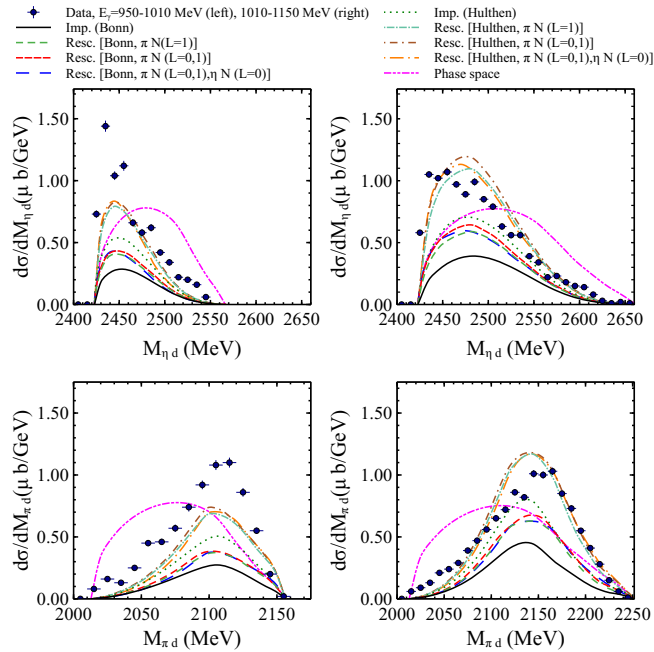}
\caption{Differential cross sections as a function of the $\eta d$ (upper panels) and $\pi^0 d$ (lower panels) invariant masses, as obtained in the impulse approximation and by considering the rescattering of $\pi$ in p-wave (orbital angular momentum $L=1$), as well as in s-wave ($L=0$), and the rescattering of $\eta$ in s-wave ($L=0$). The left (right) side figures represent average cross sections for the beam energy range $E_\gamma=950-1010$ MeV ($E_\gamma=950-1010$ MeV). The experimental data, shown as filled circles, are taken from Ref.~\cite{Ishikawa:2021yyz}.}\label{resall}
\end{figure}

Being said this, it is important to know if the rescattering mechanisms shown in Figs.~\ref{piresc}, \ref{etares} are relevant for describing the data on $\gamma d\to \pi^0\eta d$. We show the $\eta d$ and $\pi^0d$ distributions in Fig.~\ref{resall} obtained with the Bonn~\cite{Machleidt:2000ge} and Hulth\'en models~\cite{Adler:1975ga} for the deuteron wave function to illustrate the uncertainty related to the choice of the deuteron wave function parameterization.
As can be seen in Fig.~\ref{resall}, independently of the parameterization of the deuteron wave function, the effect of rescattering is relevant and leads to an increase of the strength of the mass distribution of about 50$\%$. We also find that the rescattering of a pion in p-wave, through the mechanism $\pi N\to \Delta(1232)\to \pi N$, produces the dominant contribution. The increase of the magnitude obtained for the distributions when the rescattering is implemented can be understood from the fact that the rescattering mechanism in this case helps sharing the momentum transfer between the two nucleons of the deuteron and involves the deuteron wave function at smaller momenta, where it is bigger (see Fig.~\ref{zoom}). Note, however, than even with the increase of the magnitude produced by the rescattering, the magnitude obtained for the differential cross sections for $E_\gamma=950-1010$ MeV is still smaller than that of the experimental data.

\begin{figure}[h!]
\includegraphics[width=0.35\textwidth]{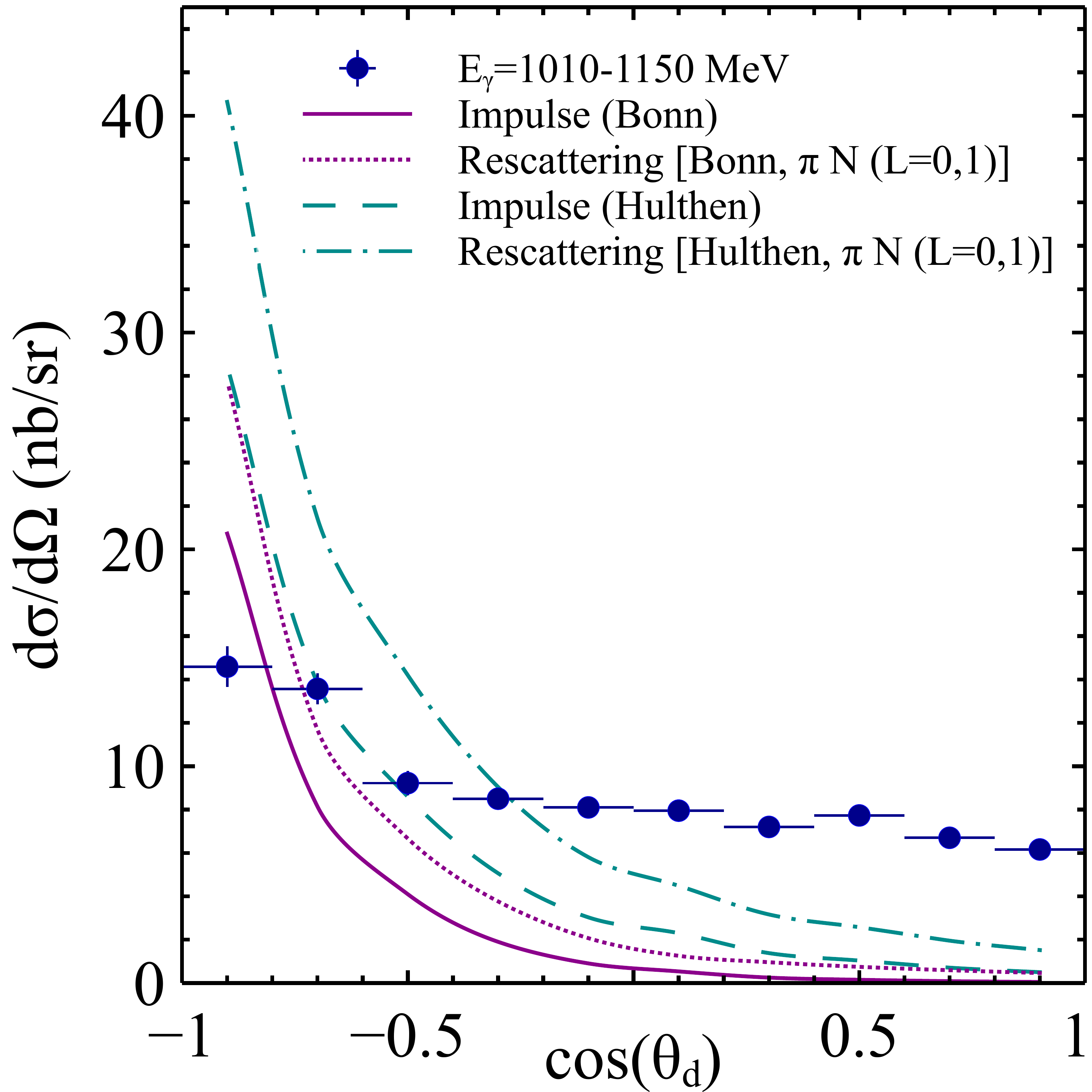}
\caption{Differential cross sections as a function of the polar angle of the outgoing deuteron. The experimental data are taken from Ref.~\cite{Ishikawa:2022mgt}.}\label{angle}
\end{figure}

To finalize this section, in Fig.~\ref{angle}  we show the results obtained on the angular distributions with the impulse approximation and with the inclusion of the rescattering processes. Since, as can be seen in Fig.~\ref{resall}, the contribution from the $\eta$ rescattering in s-wave is not significant, it is sufficient for comparing with the data to consider the effects from the rescattering of a pion. The uncertainties associated with the parameterizations of the deuteron wave function (based on Bonn and Hulth\'en potentials) are also shown. As can be seen in Fig.~\ref{angle}, the differential cross sections are  underestimated at the forward angles, while at backward angles are overestimated. Similar results have been found in Refs.~\cite{Egorov:2013ppa,Egorov:2020xdt}, and we do not have an explanation for such discrepancies, particularly since forward angle require large deuteron momenta, and even the large increase produced by the Hulth\'en wave function is clearly insufficient to reach the experimental values in the forward region.

\section{Conclusions}
In these proceedings, we have shown the results found for the $\eta d$ and $\pi^0 d$ mass distributions in the process $\gamma d\to \pi^0 \eta d$. Our description of the reaction is based on a realistic model for the $\gamma N\to \pi^0 \eta N$ process, where first $\gamma N$ couples to the resonance $\Delta(1700)$, which decays to $\eta\Delta(1232)$ and the subsequent decay of $\Delta(1232)$ to $\pi N$ produces the final state $\pi^0\eta d$. Once a $\pi$ and an $\eta$ are produced, we can also have the rescattering of these particles with the nucleons of the deuteron. The needed couplings to determine all these contributions, such as that of $ \Delta(1700)\to \eta \Delta(1232)$, are provided by previous theoretical studies. Thus, predictions for observables of the $\gamma d\to \pi^0\eta d$ reaction are obtained without fitting to the data. 

As we have shown, the shift of the data with respect to phase space can be explained with the above mentioned dynamics, and there is no need of considering the existence of dibaryons. Particularly relevant for describing the data is the contribution from the rescattering of a pion in p-wave, which increases the magnitude found for the differential cross sections with the amplitudes in the impulse approximation considerably, in as much as 50\%. 

We have also shown that the reaction investigated involves large momenta of the deuteron, in a region of momenta where  the nucleons inside the deuteron clearly overlap and it is difficult to give very precise values of the deuteron wave function. This is the reason why we used different parameterizations for the deuteron wave function, which helped us quantify the uncertainties of the theoretical calculation and they were found to be sizable.

With the mechanisms considered, the model predicts an angular distribution clearly peaking at backward angles. This result is in clear conflict with the experimental data, which correspond to a much flatter distribution.

\section*{Acknowledgements}
This work is partly supported by the Spanish Ministerio de Econom\'ia y Competitividad and European FEDER funds under Contracts No. PID2020-112777GB-I00, and by Generalitat Valenciana under contract
PROMETEO/2020/023. This project has received funding from the European Unions 10 Horizon 2020 research and innovation programme under grant agreement No. 824093 for
the “STRONG-2020” project. K.P.K and A.M.T gratefully acknowledge the travel support from the above mentioned projects. K.P.K and A.M.T also thank the financial support provided by Funda\c c\~ao de Amparo \`a Pesquisa do Estado de S\~ao Paulo (FAPESP), processos n${}^\circ$ 2019/17149-3 and 2019/16924-3 and the Conselho Nacional de Desenvolvimento Cient\'ifico e Tecnol\'ogico (CNPq), grants n${}^\circ$ 305526/2019-7 and 303945/2019-2, which has facilitated the numerical calculations presented in this work.

 \bibliographystyle{apsrev4-1}
\bibliography{bibdib}
 \end{document}